\documentclass[twocolumn,longbib]{aastex701}

\usepackage{siunitx}
\usepackage[version=4]{mhchem}
\usepackage{graphicx}
\usepackage{multirow}
\usepackage{amsmath,amssymb,amsfonts}
\usepackage{amsthm}

\usepackage[title]{appendix}
\usepackage{xcolor}
\usepackage{textcomp}
\usepackage{etoolbox}
\usepackage{natbib}

\usepackage{booktabs}
\usepackage{algorithm}
\usepackage{algorithmicx}
\usepackage{algpseudocode}
\usepackage{listings}
\usepackage{siunitx}
\usepackage[version=4]{mhchem}
\usepackage[normalem]{ulem}
\usepackage{bibunits}

\DeclareSIUnit{\angstrom}{\textup{\AA}}
\usepackage{textcomp}
\usepackage{xcolor}
\usepackage{xspace}
\newcommand{\npFe}{np\ce{Fe^0}\xspace}

\received{2025 December 15}
\revised{2026 April 1}
\accepted{2026 April 12}
\published{2026 June 10}

\begin{document}
\doi{10.3847/PSJ/ae6074}

\title{Creation of Lunar-like Rims in Ilmenite Using Synthetic Solar Wind}

\correspondingauthor{Roshan S. Trivedi}
\email{rtrivedi9@gatech.edu}
\correspondingauthor{Phillip N. First}
\email{first@gatech.edu}
\author[orcid=0009-0002-2975-3340,gname=Roshan,sname=Trivedi]{Roshan S. Trivedi}\altaffiliation{Contributed equally.}
\affiliation{School of Physics, Georgia Institute of Technology, USA}
\email{rtrivedi9@gatech.edu}

\author[orcid=0000-0003-0869-4951,gname=Advik,sname=Vira]{Advik D. Vira}\altaffiliation{Contributed equally.}
\affiliation{School of Physics, Georgia Institute of Technology, USA}
\email{avira@gatech.edu}

\author[orcid=0000-0002-6704-1064,gname=Brant M.,sname=Jones]{Brant M. Jones}
\affiliation{School of Chemistry and Biochemistry, Georgia Institute of Technology, USA}
\email{brant.jones@chemistry.gatech.edu}

\author[orcid=0000-0003-2028-447X,gname=Katherine,sname=Burgess]{Katherine D. Burgess}
\affiliation{Material Science and Technology Division, U.S. Naval Research Laboratory, USA}
\email{katherine.d.burgess@gmail.com}

\author[orcid=0000-0002-8624-1264,sname=Huang,gname=Ziyu]{Ziyu Huang}
\affiliation{Daniel Guggenheim School of Aerospace Engineering, Georgia Institute of Technology, USA}
\email{zyuhuang@gatech.edu}

\author[orcid=0000-0002-6408-2776,sname=Liu,gname=Honglin]{Honglin Liu}
\affiliation{School of Material Science and Technology, Georgia Institute of Technology, USA}
\email{hliu686@gatech.edu}

\author[orcid=0009-0001-7548-6178,sname=Rane,gname=Pranav]{Pranav Rane}
\affiliation{School of Physics, Georgia Institute of Technology, USA}
\affiliation{Division of Engineering and Applied Science, California Institute of Technology, USA}
\email{pranavrane643@gmail.com}

\author[orcid=0000-0003-2790-7799,sname=Tian,gname=Mengkun]{Mengkun Tian}
\affiliation{Institute for Matter and Systems, Georgia Institute of Technology, USA}
\email{mtian37@gatech.edu}

\author[orcid=0000-0002-1821-5689,sname=Hirabayashi,gname=Masatoshi]{Masatoshi Hirabayashi}
\affiliation{Daniel Guggenheim School of Aerospace Engineering, Georgia Institute of Technology, USA}
\email{thirabayashi@gatech.edu}

\author[orcid=0000-0002-2422-4506,sname=Orlando,gname=Thomas]{Thomas M. Orlando}
\affiliation{School of Physics, Georgia Institute of Technology, USA}
\affiliation{School of Chemistry and Biochemistry, Georgia Institute of Technology, USA}
\email{thomas.orlando@chemistry.gatech.edu}

\author[orcid=0000-0001-9884-3337,sname=Jiang,gname=Zhigang]{Zhigang Jiang}
\affiliation{School of Physics, Georgia Institute of Technology, USA}
\email{zhigang.jiang@physics.gatech.edu}

\author[orcid=0000-0003-0819-9598,sname=First,gname=Phillip]{Phillip N. First}
\affiliation{School of Physics, Georgia Institute of Technology, USA}
\email{first@gatech.edu}

\begin{abstract}

Space weathering of lunar minerals, due to bombardment from solar wind (SW) particles and micrometeoroid impacts, modifies the mineralogy within tens of nanometers of the surface, i.e., the rim. Spectroscopic signatures of these modifications, observed via remote sensing, have long been used to gauge surface exposure times on the Moon. However, the relative contributions of SW and micrometeoroids in the creation of rim features are still debated, particularly for the nanometer-scale clusters known as nanophase iron (\npFe), which commonly form in ferrous minerals. We address this issue in the laboratory, using deuterium ions and low-energy electrons as a synthetic solar wind plasma to irradiate ilmenite (\ce{FeTiO3}), a common lunar mineral. Characterization by high-resolution scanning transmission electron microscopy and electron energy-loss spectroscopy shows that the SW alone creates rims with all the main characteristics of lunar samples. We conclusively identify \npFe and quantify its distribution as a function of depth and fluence, allowing us to estimate the SW exposure of Apollo soil 71501. Our results confirm that small \npFe particles ($\SI{<10}{nm}$ in diameter) form from SW irradiation. Such experiments provide microscopic details of space weathering, improving the link between surface modification processes and macroscopic remote-sensing data.

\end{abstract}


\makeatletter
\providecommand{\@journalinfo}{}
\makeatother

\section{Introduction}
\label{sec1}

\begin{figure*} 
\centering
\includegraphics[width=\textwidth]{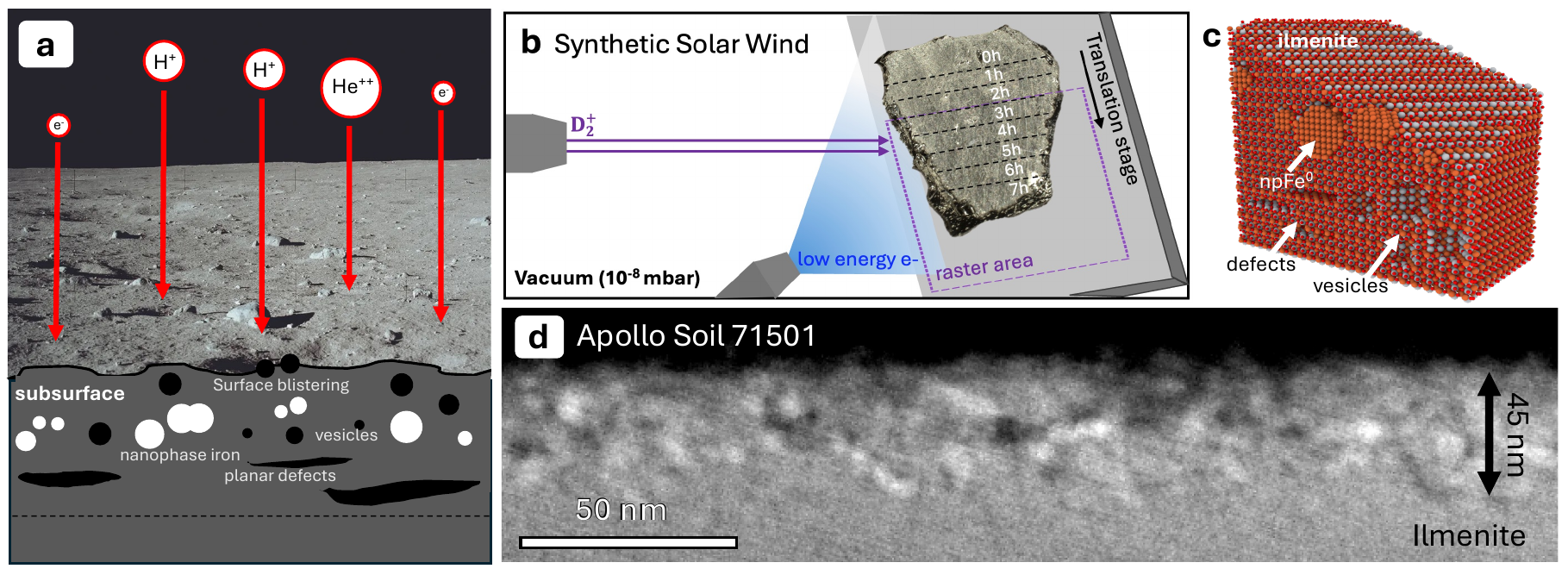}
\caption{Characteristic space weathering in lunar regolith. (a) Schematic showing SW implantation and surface alteration in lunar regolith. (b) Diagram of laboratory SSW generation using \ce{D2^+} ($\SI{1}{keV}$ per D atom) with low-energy electrons for charge balancing in the vacuum chamber (see Section~\ref{subsec:SSW}).
Ilmenite is mounted on a translation stage for a series of exposures (Table~\ref{tab:exposures}). (c) Atomic-scale model illustrating the incorporation of \npFe, planar defects, and vesicles within an ilmenite structure. (d) HAADF image showing an example of a space-weathered rim in ilmenite from Apollo soil 71501.}
\label{fig:lunar_character}
\end{figure*}

Earth’s atmosphere and magnetosphere provide effective shielding against electromagnetic radiation, atomic particles, extraterrestrial dust and even man-made space debris. In the absence of such shielding, airless bodies such as the Moon are directly exposed to the harsh space environment, which gradually alters their surfaces through a process known as space weathering. This process is broadly divided into two categories: (i) impacts of small dust particles from space (i.e., micrometeoroids) and (ii) irradiation by cosmic and solar particles (i.e., solar energetic particles and the solar wind) \citep{Pieters_space_weathering_2016,Denevi_weathering_2023}.
Space weathering modifies the micro- and nanostructure of the near-surface region (the rim) of exposed minerals, altering the optical signals observed during remote sensing \citep{Noble_2001_optical,Noble_2007_experimental,Bennett_2013_Lab_review,Grice_2025_nanoscale}. An important example is the spectral reddening and darkening observed in space-weathered lunar minerals, generally attributed to the presence of nanophase iron (\npFe) \citep{Pieters_space_weathering_2000,Keller_2001_npFe}.

Decades-long questions surround the formation mechanisms of \npFe and other characteristic features within lunar minerals \citep{Christofferson_1996_ilmenite,Pieters_space_weathering_2016,Denevi_weathering_2023,Xiong_2024_formation,Cao_rim_nature_2025}. Two space-weathering factors have been linked to the formation of \npFe: solar wind (SW; see Figure~\ref{fig:lunar_character}a) and micrometeoroid impacts \citep{Denevi_weathering_2023,Sorokin_2023_reduced,Shen_2024_separate,Xiong_2024_formation}. Based on results from returned lunar samples, the SW is thought to play a critical role in forming small \npFe particles ($\SI{<10}{nm}$), whereas larger \npFe particles ($\SI{>50}{nm}$) may require micrometeoroid impacts \citep{Shen_2024_separate}. Laboratory results are varied, however, with some concluding that micrometeoroid impacts (or large thermal events) are essential for the formation of \npFe \citep{Sasaki_2001_production,Sasaki_2003_laboratory,Loeffler_2008_laboratory,Weber_2020_laser}, while others find evidence that SW alone may be sufficient \citep{Dukes_1999_surface,Loeffler_2009_irradiation,Kuhlman_2015_simulation,Rout_2025_He,Zhang_2025_H+}. Isolating the relative contributions of these factors across a range of lunar-relevant minerals is therefore important for a fundamental understanding of space-weathering processes, as well as for assessing the potential utilization of lunar soils at proposed lunar bases.

In this paper, we focus on the effects of SW plasma in the absence of micrometeoroid impacts by performing laboratory irradiation of ilmenite (\ce{FeTiO3}), a mineral that can comprise 15\%--20\% by volume of \ce{Ti}-rich mare basalts \citep{Sourcebook_1991_minerals}. Ilmenite has also been linked to the lunar water cycle \citep{Xu_widespread_2025}, making it crucial to understand its interaction with the SW plasma. Our laboratory instrumentation generates synthetic solar wind (SSW; see Figure~\ref{fig:lunar_character}b) using ionized deuterium (\ce{D2}) molecules ($\SI{1}{keV}$ per D atom) and low-energy electrons ($\SI{20}{eV}$). We expose terrestrial ilmenite to lunar-equivalent fluences corresponding to $\sim$200 to $\sim$1700 yr of exposure at the equator, and employ scanning transmission electron microscopy (STEM) with electron energy-loss spectroscopy (EELS) to systematically characterize the evolution of \npFe with SSW exposure. These measurements demonstrate that SW alone is sufficient to create space-weathered rims, rich with vesicles and \npFe, similar to those seen in lunar ilmenite-bearing soils (see Figures~\ref{fig:lunar_character}c,d) \citep{BURGESS_2018_ilmenite,Guo_Zhang_Ti_ilmenite_2024,Zhang_ilmenite_2024,Cao_rim_nature_2025}. Our detailed characterization of the fluence-dependent \npFe density distribution reveals systematics that provide an independent measure of the exposure age of lunar regolith samples.

\section{Methods}\label{sec:methods}

\subsection{Synthetic Solar Wind: Laboratory Irradiation}\label{subsec:SSW}

The SSW irradiation was generated in a vacuum chamber with a base pressure of 10$^{-8}$ mbar. The chamber was flooded with deuterium (\ce{D2}) molecules to a pressure of 10$^{-6}$ mbar to produce the SSW using a sputter gun (Physical Electronics Industries, model $\#$20-115). The ions were accelerated by a potential difference of $\SI{2}{kV}$ between the filament and the sample. Upon impact, the \ce{D2^+} ions dissociate into a $\SI{1}{keV}$ \ce{D} atom and a $\SI{1}{keV}$ \ce{D+} ion (as described in \citealt{SIGMUND_1996_collision}). The beam has a full width at half-maximum (FWHM) of $\SI{2}{mm}$, which was rastered over an area of $8\times8$~$\SI{}{mm^2}$. The beam moves across the fast (slow) axis at a rate of $\SI{7500}{Hz}$ ($\SI{500}{Hz}$), completing one cycle in $\SI{0.002}{s}$. The beam current in the sputter gun was measured through a circular Faraday cup (Kimball Physics FC-70) with an aperture area of $\SI{2}{mm^2}$. A $\SI{20}{eV}$ electron flood gun (Surface Science Laboratories, 8711 charge neutralizer) was used for charge passivation of the mineral surface. Both the sputter and electron beams were carefully aligned with the sample to expose the desired region. The sample was moved laterally using a translation stage through the beam in $\SI{1}{mm}$ increments to enable multiple exposures within a given sample.

\subsubsection{Ilmenite Preparation}

The terrestrial ilmenite sample was obtained from the Georgia Tech Earth and Atmospheric Sciences teaching collection. Its composition was verified using X-ray fluorescence (Bruker M4 Tornado, available through the Georgia Tech Materials Characterization Facility). The sample surface was polished with 120-grit silicon carbide sandpaper on a lapping machine (South Bay Technology, model 910). A diamond scriber was used to mark the exposure region into 1~mm-wide sections for easy identification during the multi-exposure process.

Based on high-resolution STEM measurements, we found that the ilmenite sample is polycrystalline and contains dispersed small hematite inclusions (identified from $d$-spacing). This assessment was further supported by elemental composition analysis using energy-dispersive spectroscopy and is consistent with the presence of \ce{Fe^{3+}} found in EELS measurements.

\subsubsection{Deuterium Exposure}

All SSW exposures in this study were conducted at an average \ce{D^+} flux of $\SI{3E14}{cm^{-2} s^{-1}}$ (corresponding to a beam current of $\SI{500}{nA}$) with an uncertainty of $\pm5\%$. Two series of exposures were performed to achieve a maximum fluence of $\sim1700$~lunar-equivalent years.
During irradiation, the sample was moved laterally into the beam in $\SI{1}{mm}$ increments, as shown schematically in Figure~\ref{fig:lunar_character}b. Due to beam size constraints, the two series were conducted on separate samples: one with a total exposure time of $\SI{7}{hr}$ (sample moved every hour) and the other with a total time of $\SI{14}{hr}$ (sample moved every $\SI{2}{hr}$). Exposure fluences and the corresponding lunar-equivalent time at the lunar equator---calculated using an average SW \ce{H^+} flux of $\SI{3E8}{cm^{-2}s^{-1}}$ \citep{COllier_2014_solar_wind}---for the lift-out samples studied in this work are summarized in Table \ref{tab:exposures}.
\begin{table}[h!]
    \centering
    \caption{Irradiation parameters of terrestrial ilmenite.}
    \label{tab:exposures}
    \begin{tabular}{|c|c|c|}
        \hline
        Time & Fluence & Lunar-equivalent Time \\
        (hr) & $(\SI{E18}{D^{+}\per cm^{2}})$ & (yr) \\
        \hline
        2   & 2.2  & 238  \\
3   & 3.4  & 356  \\

5   & 5.6  & 594  \\
7   & 7.9  & 832  \\
8   & 9.0  & 950  \\
12  & 13.5 & 1426 \\
14  & 15.7 & 1662 \\
        \hline
    \end{tabular}
\end{table}
\subsection{Electron Microscopy}

\subsubsection{Lift-out Sample Preparation}\label{subsubsec:liftouts}

A focused ion beam (FIB) was used to lift out different regions of the irradiated ilmenite with a Thermo Fisher Helios 5CX FIB-SEM at the Georgia Tech Materials Characterization Facility. The instrument produces \ce{Ga^+} ions at 0.5--30 keV (currents between $\SI{1}{pA}$ and $\SI{100}{nA}$) and electrons at 0.5--30 keV (maximum beam current $\SI{176}{nA}$). Exposed sample surfaces were coated with $\SI{20}{nm}$ of amorphous carbon and imaged at $\SI{5}{keV}$ using electrons to assess irradiation-induced damage. Following scanning electron microscopy (SEM) imaging, an additional $\SI{40}{nm}$ of Au-Pd coating was deposited to further protect the surface during FIB milling. A standard lift-out procedure (e.g., \citealt{tomus_situ_2013}) was used to extract a $2\times5~\SI{}{\mu m^2}$ lamella, which was then polished to $\SI{<50}{nm}$ thickness for high-resolution microscopy.

\subsubsection{STEM Imaging}\label{subsubsec: STEM}

The lift-out samples were imaged using a Hitachi HD-2700 STEM at the Georgia Tech Materials Characterization Facility. The microscope was operated at $\SI{200}{keV}$ with a convergence angle of $\SI{27}{mrad}$ and a spatial resolution of $\Delta r\simeq1.3$~\AA. High-angle annular dark-field (HAADF) images were collected to characterize the SSW rims. Large-scale HAADF images ($\sim 200\times\SI{200}{nm^2}$) were used to quantify the nanophase iron (\npFe) distribution and were stitched in Adobe Illustrator to produce panoramic images.

Image segmentation was used to determine the size of individual \npFe regions and their depth-dependent distribution, $\chi(z)$, from the surface. We selected \npFe for segmentation because it has a higher contrast in HAADF images (compared to vesicular and \ce{Ti}-rich regions), making it a feature whose distribution can be reliably quantified from STEM results. Preprocessing was done in Gwyddion \citep{gwyddion}, an open-source software for analysis of Scanning Probe Microscopy data, to mask the non-\npFe regions via thresholding \citep{Vincent_1991_segmentation}. Segmentation was performed using AnyLabeling, an open-source, artificial intelligence (AI)-powered computer vision tool.
Since built-in AI models (e.g., the Segment Anything Model, SAM) performed poorly on our images, \npFe boundaries were delineated manually based on empirical observation and human visual judgment, guided by HAADF grayscale variations. Annotation assumed smooth, continuous phase interfaces, consistent with minimal interfacial energy. The resulting JSON files were used to extract $\chi(z)$, the peak value $\chi^{\mathrm{max}}$, and the integrated density $I_\chi=\int \chi(z)dz$ for each lift-out sample. $\chi(z)$ plots used a bin width of $\SI{5}{nm}$, matching the average surface height variation. The integrated density $I_\chi$ is normalized by a constant rim width of $\SI{60}{nm}$ for the power-law fit.

\subsubsection{UltraSTEM/EELS}\label{subsubsec:EELS}

EELS measurements were performed at the Naval Research Laboratory using a Nion UltraSTEM200-X equipped with a Gatan Enfinium ER spectrometer.
The STEM/EELS was operated at $\SI{200}{keV}$ and $\SI{40}{pA}$, with a $\SI{0.1}{nm}$ probe and a $\SI{27}{mRad}$ convergence angle. EELS core-loss and low-loss spectra were acquired sequentially with a dispersion setting of $\SI{0.1}{eV}$ per channel for each element: \ce{Ti} (starting core-loss spectra at $\SI{435}{eV}$), $\text{O}$ ($\SI{510}{eV}$), and \ce{Fe} ($\SI{687}{eV}$). This setting enabled the capture of fine-structure details for each element, while also including the \ce{Ti} and \ce{Fe} M-edges together in the low-loss spectra.

To create the oxygen and iron concentration maps, separate power laws were first fitted to and subtracted from the \ce{O} K-edge and \ce{Fe} L$_{2,3}$-edges of the core-loss spectra to remove background contributions. The resulting background-subtracted spectra were then integrated over their respective edges to map the relative concentrations of \ce{O} and \ce{Fe} (see Appendix~\ref{app:SI-figures}; Figures~\ref{fig:SI-EELS_raw} and \ref{fig:SI-Fe_O_conc}). Maps of the \ce{Fe}/\ce{Ti} ratio were generated using the M$_{2,3}$-edges by subtracting a common power-law background from both edges and integrating over each. The ratio maps were used to mask regions with high \ce{Ti} and high \ce{Fe} concentrations. The masked low-loss spectra were plotted alongside a reference spectrum from an area unaffected by SSW irradiation ($z>\SI{55}{nm}$) to highlight SSW-induced changes. For the \ce{Ti} L$_{2,3}$ edges, the power-law background was subtracted after averaging over the masked regions due to the low signal-to-noise ratio. A similar analysis procedure was used for the \npFe regions and \ce{Fe} L$_{2,3}$-edges.

\subsection{Numerical Simulations}

\subsubsection{MC Calculations}\label{subsubsec:SRIM}

Monte Carlo (MC) simulations were executed to investigate the ion-solid interactions and potential defect-forming mechanisms behind our laboratory irradiation results. MC codes use a binary collision approximation to model the collision cascade, where interactions are treated as a sequence of elastic (nuclear), ballistic collisions governed by screened Coulombic interatomic potentials e.g., Kr--C \citep{wilson_1977} or Ziegler--Biersack--Littmark (ZBL; \citealt{BIERSACK1980257}). Alongside this, inelastic (electronic) energy loss is treated as a friction-like interaction as ions traverse an electron gas \citep{Lindhard_1961, Oen_1976}; for low-energy light ions (e.g., $\SI{1}{keV}$ \ce{D^+}), this results in electronic stopping that increases approximately proportional to ion velocity. We use the Transport of Ions in Matter (TRIM; \citealt{srim_textbook,ziegler1977stopping}) module within the Stopping and Range of Ions in Matter (SRIM) code to probe defect-forming mechanisms in ilmenite.  TRIM is a powerful and widely used tool for modeling ion–matter interactions across more than 6 orders of magnitude in energy, but its generality introduces limitations in specific applications \citep{WITTMAACK201657}. SDTrimSP was introduced to address some of the limitations of TRIM, especially for modeling ion bombardment with low-energy ions ($<\SI{1}{keV}$) relevant to space-weathering modifications of planetary materials (see Appendix~\ref{app:SDTrimSP}) \citep{Wittmaack_2017, SZABO202247, Morrissey_2023, mutzke2024sdtrimsp}. We utilize both MC codes (SRIM/TRIM and SDTrimSP) to compare with our experimental results.

We run TRIM simulations using pySRIM \citep{Ostrouchov2018} for 1 million incident $\SI{1}{keV}$ \ce{D^+} ($\SI{2}{amu}$ hydrogen) ions interacting with a $\SI{60}{nm}$ thick ilmenite layer (density of $\SI{4.78}{g/cm^3}$).
We used the full cascade collision mode and obtained the energy partitioning into electronic and nuclear losses from the IONIZ and E2RECOIL files, respectively, and implantation profiles from the RANGE file \citep{srim_documentation}. To compare TRIM with SDTrimSP, we run SDTrimSP simulations in static mode for an ilmenite target (thickness of $\SI{60}{nm}$, composed of \ce{FeO} and \ce{TiO2}) for 1 million \ce{D^+} ions at $\SI{1}{keV}$. We employ the Ziegler-Biersack-Littmark (ZBL)
potential \citep{wilson_1977,Wittmaack_2017, mutzke2024sdtrimsp}, using Gauss–Legendre integration and the Lindhard–Scharff inelastic loss model for all calculations \citep{Lindhard_1961}.
The depth distributions of energy partitioning and damage profiles are saved using the \texttt{lenergy\_distr} flag. We provide a comparison between TRIM and SDTrimSP in Appendix~\ref{app:SDTrimSP}. Given the agreement between the codes, we display the SDTrimSP profiles in Section~\ref{subsubsec:defect_production}, as it uses reliable stopping powers in the $\SI{}{keV}$ energy range \citep{Wittmaack_2017}.

\subsubsection{Density Functional Theory Calculations}

First-principles calculations for \ce{FeTiO3} were performed using the Quantum ESPRESSO package \citep{QE_ref}, based on density functional theory (DFT; \citealt{HohenbergKohn_ref, KohnSham_ref}). The initial lattice parameters were set as $a=5.53937$~\AA\ and $\cos(\gamma)=0.578097$, corresponding to hexagonal symmetry with a unit cell containing 19 atoms, comprising \ce{Fe}, \ce{Ti}, and \ce{O} species. Because all oxygen atoms in the unit cell are equivalent, a single oxygen atom at random was removed to represent the defect created by irradiation. Exchange-correlation effects were treated using the Perdew--Burke--Ernzerhof functional within the generalized gradient approximation. A plane-wave basis set with a kinetic energy cutoff of 60~Ry was used for the electronic wave functions, while the charge density cutoff was set to 720~Ry. Gaussian smearing with a broadening of 0.1~Ry was applied to the electronic occupations to aid convergence. Spin-polarized calculations were enabled, with initial magnetizations of 0.2~Bohr magnetons assigned to \ce{Fe} and \ce{Ti} atoms, while oxygen atoms were initialized with zero magnetization. Our results in Figure~\ref{fig:SI-dft} indicate that the total energy stabilizes at \texttt{ecutwfc} = 60~Ry, demonstrating that this value is sufficient for reliable convergence in our simulation.

\section{Results}\label{sec2}
\subsection{Characteristics of Weathered Rims in Lunar Regolith}

\begin{figure*}[!tbp]
\centering
\includegraphics[width=\textwidth]{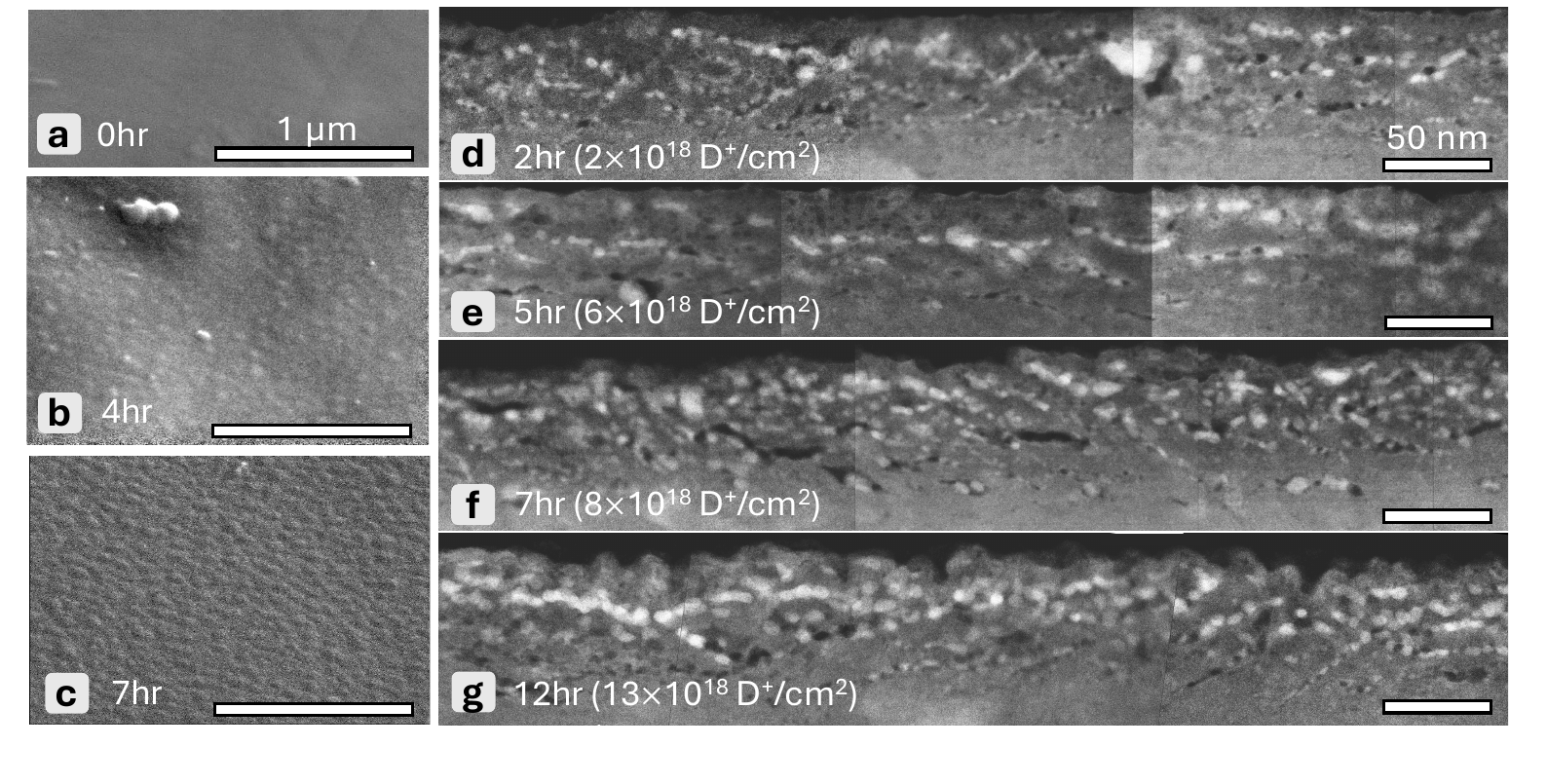}
\caption{SSW-induced surface alteration in terrestrial ilmenite. (a)--(c) SEM images of ilmenite showing SSW-induced surface damage after exposures of $\SI{0}{hr}$ (a), $\SI{4}{hr}$ (b), and $\SI{7}{hr}$ (c). (d)--(g) HAADF cross-sectional images of ilmenite rims irradiated for $\SI{2}{hr}$ (d), $\SI{5}{hr}$ (e), $\SI{7}{hr}$ (f), and $\SI{12}{hr}$ (g). Corresponding fluence and lunar-equivalent years are shown in Table~\ref{tab:exposures}. Bright spots indicate \npFe, while dark regions correspond to vesicles and/or planar defects.}
\label{fig:blisters_rims}
\end{figure*}

\begin{figure*} 
\centering
\includegraphics[width=\textwidth]{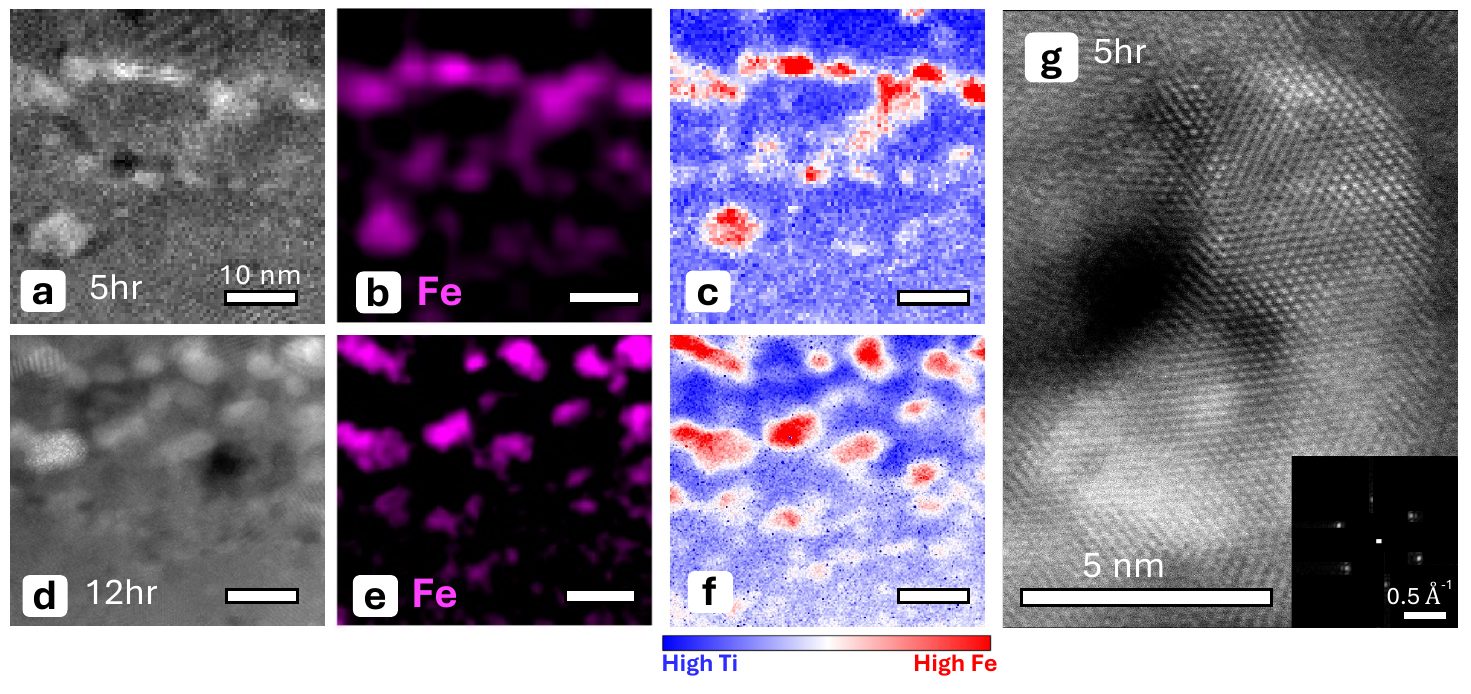}
\caption{Compositional and structural characterization of \npFe regions in SSW rims. (a) HAADF image of the $\SI{5}{hr}$ sample. (b) \ce{Fe} concentration map (see Section~\ref{subsubsec:EELS}) of the $\SI{5}{hr}$ sample in the same region as (a). (c) \ce{Fe}/\ce{Ti} ratio map (see Section~\ref{subsubsec:EELS}) of the $\SI{5}{hr}$ sample in the same region as (a). (d)--(f) Same as (a)--(c), but for the $\SI{12}{hr}$ sample. (g) High-resolution HAADF image of \npFe in the $\SI{5}{hr}$ sample with $d$-spacing corresponding to the \{110\} family of planes in $\alpha$-Fe. Inset: fast Fourier transform (FFT) of the crystalline \npFe region.}
\label{fig:npFe_summary}
\end{figure*}

For comparison with laboratory SSW experiments, we carried out a STEM study of space-weathered lunar ilmenite from Apollo soil 71501. Figure~\ref{fig:lunar_character}d shows an example HAADF image of the space-weathering-induced features, where the dark spots indicate vesicular regions and the bright spots correspond to \npFe. Due to its high Fe content and greater resistance to amorphization compared to silicates, ilmenite often retains partially crystalline regions in space-weathered rims (i.e., antiamorphous regions) \citep{Cao_rim_nature_2025}. Within lunar rims, \npFe, vesicles, and planar defects are commonly observed, even where the surrounding ilmenite is not fully amorphized \citep{Zhang_ilmenite_2024,Guo_Zhang_Ti_ilmenite_2024}.
Our systematic laboratory exposures allow us to isolate the effects of SW in creating these rims and reveal a fluence dependence of rim characteristics. This fluence dependence provides a new independent measure of exposure age using the \npFe distribution within the SSW rims of terrestrial ilmenite. We then apply our \npFe-based calibration to ilmenite grains from 71501 to estimate their SW exposure age (see Section~\ref{subsubsec:Density}).

\subsection{Synthetic Solar Wind Rims in Ilmenite}

SEM images of SSW-exposed ilmenite reveal surface alteration induced by the irradiation (Figures~\ref{fig:blisters_rims}a--c), with the density of surface damage increasing with exposure time. The postexposure surface texture is consistent with previous reports \citep{Gu_2022_blistering,Cymes_2024_helium}, which attribute the surface alterations or blistering to gas buildup.

HAADF images from FIB lift-out samples of SSW-exposed ilmenite (see Section~\ref{subsubsec: STEM}) are shown in Figure~\ref{fig:blisters_rims} for different exposure times of $\SI{2}{hr}$, $\SI{5}{hr}$, $\SI{7}{hr}$, and $\SI{12}{hr}$. These SSW-induced rims in terrestrial ilmenite exhibit characteristics similar to those of space-weathered regolith (e.g., Figure~\ref{fig:lunar_character}d), indicating that the laboratory SSW flux and exposure times provide a reasonable simulation of lunar SW conditions.
Within the top $\SI{60}{nm}$ of the SSW rims (e.g., Figure~\ref{fig:blisters_rims}d), there are bright regions that we identify as \npFe (based on results from Section~\ref{subsec:npFe_characterization}), vesicles (dark circular regions), and elongated planar defects (dark elongated regions). We frequently observe clustering of \npFe around vesicular regions and/or planar defects (e.g., right side of Figure~\ref{fig:blisters_rims}g).

\subsection{\texorpdfstring{Structure of np\ce{Fe^0} and Surrounding Regions}{Structure of nanophase iron and Surrounding Regions}}\label{subsec:npFe_characterization}

Figures~\ref{fig:npFe_summary}a and (d) show HAADF images of the SSW-exposed rims for the $\SI{2}{hr}$ and $\SI{5}{hr}$ samples. Figures~\ref{fig:npFe_summary}b and (e) highlight the \ce{Fe}-rich regions throughout the SSW rim, which correspond to the bright (\npFe) regions in the HAADF images (Figures~\ref{fig:npFe_summary}a,d). An anticorrelation between \ce{O} and \ce{Fe} concentrations in the SSW rims (Figure~\ref{fig:SI-Fe_O_conc}) indicates that the \ce{Fe}-rich regions are relatively oxygen-free, suggesting the presence of metallic iron. The \ce{Fe}/\ce{Ti} ratio maps (Section~\ref{subsubsec:EELS}) of the same regions are shown in Figures~\ref{fig:npFe_summary}c and f. Moreover, a high-resolution HAADF image of a selected \ce{Fe}-rich region is shown in Figure~\ref{fig:npFe_summary}g, where the atomic structure of \npFe is clearly resolved and identified as body-centered cubic (bcc). The $d$-spacing is measured to be 2.02~\AA, corresponding to the \{110\} family of lattice planes in $\alpha$-Fe.
Within the SSW rims, ilmenite lattice fringes are still visible in STEM (Figure~\ref{fig:SI-npFe_spacing}), indicating that the sample is not fully amorphized, consistent with observations in lunar ilmenite and olivine \citep{Cao_rim_nature_2025}.

\begin{figure*}
\centering
\includegraphics[width=0.95\textwidth]{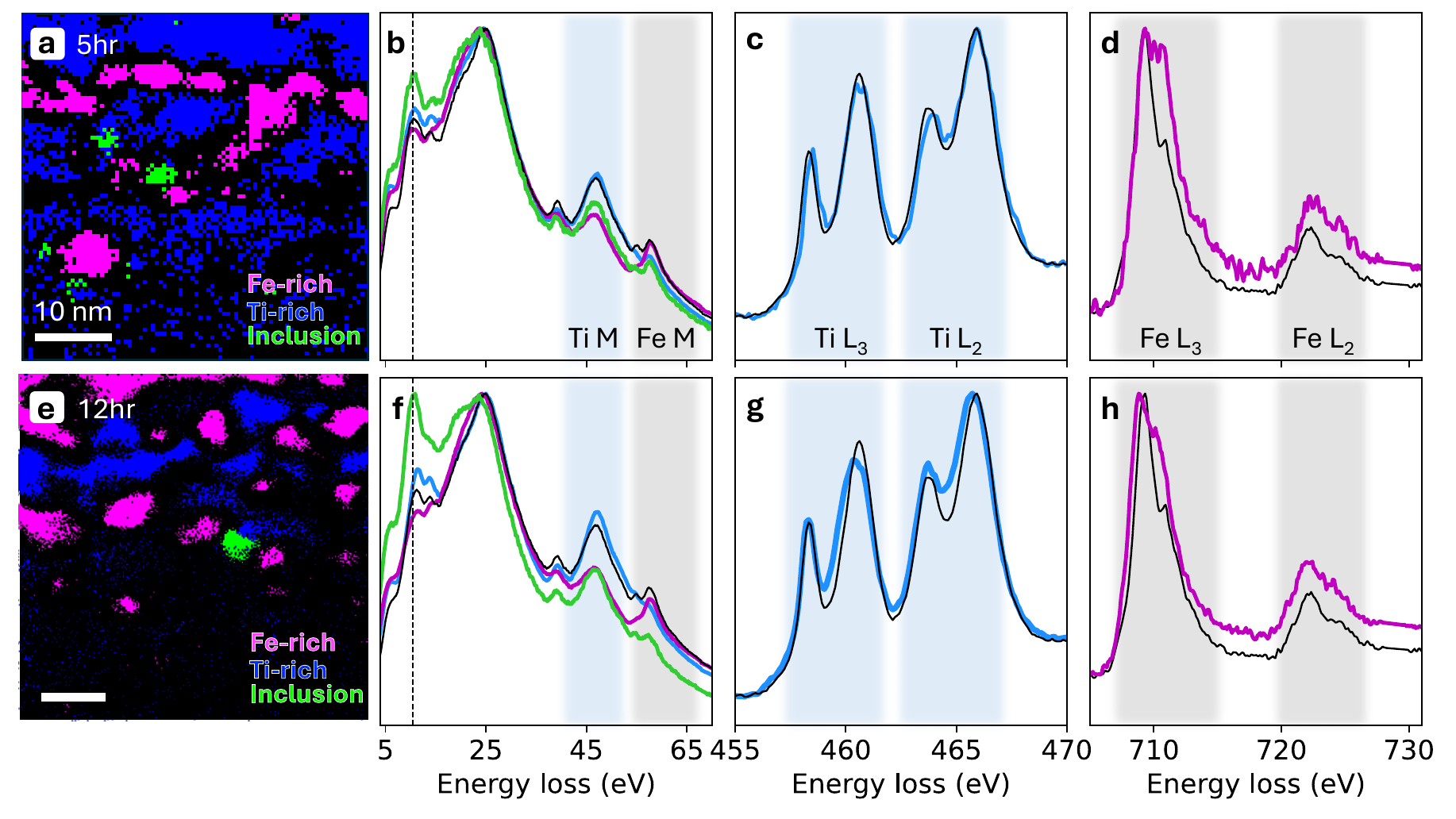}
\caption{EELS fine structure of features in SSW rims. (a) Overlaid \ce{Fe} (magenta) and \ce{Ti} (blue) concentration maps (see Section~\ref{subsubsec:EELS}) of the $\SI{5}{hr}$ sample in the same region as Figure~\ref{fig:npFe_summary}a. The dark inclusions in Figure~\ref{fig:npFe_summary}a are shown in green. (b) EELS low-loss spectra averaged over the color-coded regions in (a). The dashed line indicates a characteristic $\SI{\sim10.5}{eV}$ peak. (c) \ce{Ti} L-edge ELNES spectra from the \ce{Ti}-rich (blue) region in (a). (d) \ce{Fe} L-edge ELNES spectra from the \ce{Fe}-rich (magenta) region in (a). All spectra are plotted alongside that from unexposed ilmenite (black). (e)--(h) Same as (a-d), but for the $\SI{12}{hr}$ sample.}
\label{fig:inclusion_spectra}
\end{figure*}

The oxidation state of \npFe in the SSW rims can be confidently identified using electron energy-loss near-edge structure (ELNES). We find that \npFe exhibits distinct ELNES spectra at the $\text{Fe}\ \text{L}_{2,3}$-edges compared to terrestrial ilmenite, which contains predominantly \ce{Fe^{2+}} and minor \ce{Fe^{3+}} \citep{Gu_Lin_ferric_2023}. Figures~\ref{fig:inclusion_spectra}d and h show a comparison of \ce{Fe} L-edge ELNES spectra between unexposed ilmenite (black) and \npFe (magenta), averaged over the Fe-rich regions in Figures~\ref{fig:inclusion_spectra}a and e, corresponding to the same regions in Figures~\ref{fig:npFe_summary}b and e. While a small difference in the onset energy of the L$_{2,3}$ edges between \ce{Fe^0} and \ce{Fe^{2+/3+}} has previously been reported \citep{Gu_Lin_ferric_2023}, such a shift is difficult to resolve in our data (Figures~\ref{fig:inclusion_spectra}d,h) due to the limited averaging time necessary to avoid beam damage in the lift-out samples. Instead, our data show that \npFe exhibits a higher continuum than oxidized \ce{Fe} in ilmenite (Figures~\ref{fig:inclusion_spectra}d,h), a trend well established in the literature for metallic iron \citep{Garvie_1998_Ferrous,Feldhoff_2009_Spin_iron}. Thus, based on our STEM/EELS measurements (bcc structure of $\alpha$-Fe in Figure~\ref{fig:npFe_summary}g, \ce{O-Fe} anticorrelation in Figure~\ref{fig:SI-Fe_O_conc}, and higher continuum in Figures~\ref{fig:inclusion_spectra}d, (h)), we conclusively demonstrate that \npFe can form from SSW alone.

With the formation of \npFe comes another question: What happens within the now iron-depleted regions surrounding \npFe? Based on the intensity ratio of \ce{Fe} and \ce{Ti} \ce{M}-edges, we observe \ce{Ti}-rich regions (blue in Figures~\ref{fig:npFe_summary}c and f) that surround the \npFe (red in Figures~\ref{fig:npFe_summary}c and f). Within the \ce{Ti}-rich regions, the \ce{Ti} \ce{L}-edge ELNES spectra differ slightly from unexposed ilmenite, particularly with prolonged SSW exposure (Figures~\ref{fig:inclusion_spectra}c,g and Figure~\ref{fig:SI-Ti-richregions}).
From the HAADF images (Figures~\ref{fig:npFe_summary}a,d), we observe a few circular dark inclusions within the \ce{Ti}-rich regions. The appearance of these features is similar to vesicles, but the EELS low-loss spectra over these dark inclusions reveal a $\SI{\sim10.5}{eV}$ peak (Figure~\ref{fig:inclusion_spectra}b), becoming more pronounced with prolonged SSW exposure (Figure~\ref{fig:inclusion_spectra}f). A similar $\sim\SI{10.5}{eV}$ peak was observed in space-weathered lunar ilmenite \citep{BURGESS_2018_ilmenite} and attributed to the formation of \ce{TiO2}, suggesting that the spectral alteration in the inclusions within our sample is due to the formation of a \ce{TiO2} phase \citep{Launay_2004_TiO2}. A detailed comparison of our \ce{Ti} ELNES spectra in \ce{Ti}-rich regions to rutile and anatase phases of \ce{TiO_2} (Figure~\ref{fig:SI-cascade_spectra}) shows similarities in the edge features, although the spectral differences we observe are not sufficient to determine the specific phase. Measurements of SW-exposed lunar ilmenites by others (e.g., \citealt{BURGESS_2018_ilmenite,Zhang_ilmenite_2024}) produced comparable results.

\section{Discussion}\label{sec12}

Our laboratory investigations demonstrate that SSW rims in ilmenite exhibit the same complex characteristics found in returned lunar samples, including \npFe, vesicles, and \ce{Ti}-rich regions. The formation dynamics of these rim features is undoubtedly complex. Nevertheless, we focus on the most prominent feature of SSW rims,  \npFe, and apply analyses similar to prior work that revealed growth mechanisms in less complex materials \citep{Dennis_1978_model,holland_1985,Harbsmeier_1998_amorphization}.

\subsection{\texorpdfstring{\npFe Distributions}{npFe Distributions}}
Models of radiation effects commonly focus on the creation and growth of amorphous regions within the material \citep{Gibbons_1972,Harbsmeier_1998_amorphization}. The formation of \npFe and accompanying rim features is somewhat different, but the initiation mechanisms may be similar.  Since the most identifiable product is \npFe (not amorphous regions), we use the density variation of \npFe with fluence and depth as the quantity of interest for comparison with existing models of radiation-induced amorphization clusters \citep{Gibbons_1972,Harbsmeier_1998_amorphization}.

\subsubsection{Density Variation with Fluence}\label{subsubsec:Density}
\begin{figure*}[!tbp]
\centering
\includegraphics[width=\textwidth]{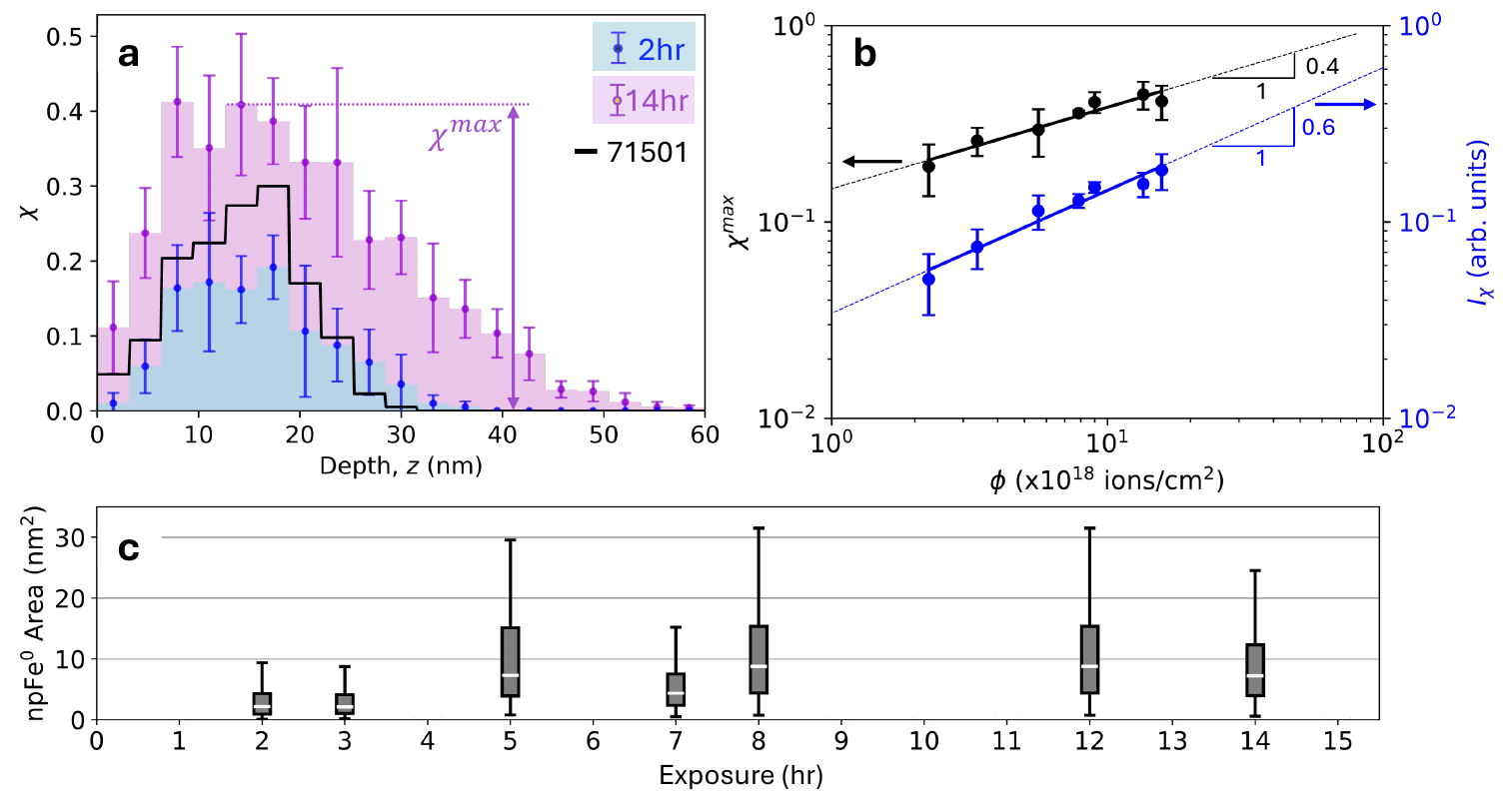}
\caption{Distribution of \npFe in SSW rims. (a) Density distribution of \npFe ($\chi$) vs. depth for the SSW rims ($\SI{2}{hr}$ in blue, $\SI{14}{hr}$ in purple), overlaid with the distribution for 71501 ilmenite (black). (b) Power-law fits to the maximum of $\chi$ distributions ($\chi^{\mathrm{max}}$; left axis, black) and normalized integrated density of \npFe coverage ($I_\chi$; right axis, blue). (c) Size distribution of \npFe for all exposures. Data for \npFe distributions are extracted from image segmentation of HAADF images (see Section~\ref{subsubsec: STEM}).}
\label{fig:npFe_distribution}
\end{figure*}

In Figure~\ref{fig:npFe_distribution}a, the density distribution of \npFe, denoted as $\chi(z)$, is derived from HAADF image segmentation and defined as the percent coverage of \npFe within $\SI{5}{nm}$ bins (see Section~\ref{subsubsec: STEM}). Both the maximum density ($\chi^{\mathrm{max}}$) and integrated density ($I_\chi$) of the \npFe distributions increase with exposure time (Figure~\ref{fig:npFe_distribution}b). The increase in $\chi^{\mathrm{max}}$ indicates that these exposures are performed in a regime before the saturation of radiation-induced damage \citep{Harbsmeier_1998_amorphization}. Amorphization models typically describe fluence dependence by an exponential saturation behavior and use a power-law distribution as a first-order approximation, which is valid over a narrow fluence regime (see Appendix \ref{app:amorphization-models}) \citep{Dennis_1978_model,Harbsmeier_1998_amorphization}. Power-law fits to $\chi^{\mathrm{max}}$ and $I_\chi(\phi)$ yield: $\chi^{\mathrm{max}}(\phi)=(\phi/\phi_c)^m$ where $\phi_c=1.0\times10^{20}\ \SI{ }{ions/cm^2}$ (representing the critical amorphization fluence; see \citealt{Harbsmeier_1998_amorphization}) and $m=0.4$, and $I_\chi(\phi)/I_\chi(\phi_c)=(\phi/\phi_c)^n$ where $n=0.6$. Fractional amorphization theory predicts $n=0.5$ (Appendix~\ref{app:amorphization-models}), implying that the physics behind homogeneous growth models aligns in part with the dose dependence of \npFe. While the appropriate model of formation dynamics is yet to be determined, the observed power-law behavior in Figure~\ref{fig:npFe_distribution}b can be used to establish an independent measure of exposure age using our \npFe distributions in SSW rims of terrestrial ilmenite.

We estimated the approximate SW exposure age of 71501 ilmenite based on the HAADF image in Figure~\ref{fig:lunar_character}d. The \npFe distributions result in a normalized integrated density $I_\chi$ of $\SI{\sim0.08}{}$ and $\chi^{\mathrm{max}}$ of $\SI{\sim0.30}{}$ (Figure~\ref{fig:npFe_distribution}a), which can be used to determine the exposure fluence (and resulting SW exposure age). The power-law fits in Figure~\ref{fig:npFe_distribution}b result in an approximate SW exposure of 450 and 670 lunar-equivalent years based on $I_\chi$ and $\chi_{max}$, respectively. Solar energetic particle (SEP) tracks for 71501 indicate an exposure age between 14 kyr and 2 Myr \citep{Keller_2021_SEP}. However, SEP tracks record the time a grain spent within the top several millimeters of the space-exposed surface, whereas the SW exposure age is the period of time the specific portion of the grain was at the exact surface and is necessarily shorter. The maturity index of 71501 is $\SI{\sim35}{}$, placing it in the submature (30-60) category \citep{Morris1976}. This is consistent with our estimated exposure corresponding to a low-intermediate fluence range (3--5 hr). The maturity index for lunar soils is based on ferromagnetic resonance and is a measure of the amount of fine-grained (superparamagnetic) metal particles (i.e., \npFe) normalized by the total amount of \ce{FeO} in the sample. Thus, the general agreement between our exposure estimate for this grain and the bulk maturity index demonstrates the potential for providing more precise exposure ages for the near-surface regions of lunar soils using our technique. Additional experiments for differing exposure times and fluxes in SSW-exposed terrestrial simulants would allow for improved accuracy in determining the SW age of returned lunar regolith using \npFe distributions.

\subsubsection{Size Variation with Fluence}

The size of \npFe particles can be indicative of different types of space weathering and is known to influence remote-sensing measurements.
For instance, large \npFe ($\SI{>50}{nm}$) contributes to the darkening of remote optical reflectance spectra, and small \npFe ($\SI{<10}{nm}$) causes reddening \citep{Noble_2007_experimental}. Here, we isolate the effects of SW-induced \npFe and track the distribution of \npFe particle size with increasing deuterium exposure (Figure~\ref{fig:npFe_distribution}c). We do not observe a significant change in the median particle size, where the median \npFe area varies between $\SI{2.1}{nm^2}$ (or $\SI{\sim1.6}{nm}$ in diameter, assuming a circular area) and $\SI{\sim8.7}{nm^2}$ (or $\SI{\sim3.4}{nm}$ in diameter). There are a few outliers that have areas of up to $\SI{\sim90}{nm^2}$ or diameters of $\SI{\sim10}{nm}$ (e.g., the right side of Figure~\ref{fig:blisters_rims}d), but it remains unclear if those are single \npFe particles or conglomerations of several. Aside from these outliers, we do not observe any systematic trend in \npFe area with exposure (Figure~\ref{fig:npFe_distribution}c), implying that substantially larger \npFe found in some regolith grains may require additional energy input for its formation. This is in agreement with \citet{Shen_2024_separate}, where it was proposed that small \npFe are produced by SW ($\SI{<10}{nm}$ in diameter), and large \npFe particles require large thermal events, such as those from micrometeoroid impacts.

\begin{figure*}[!tbp]
\centering
\includegraphics[width=\textwidth]{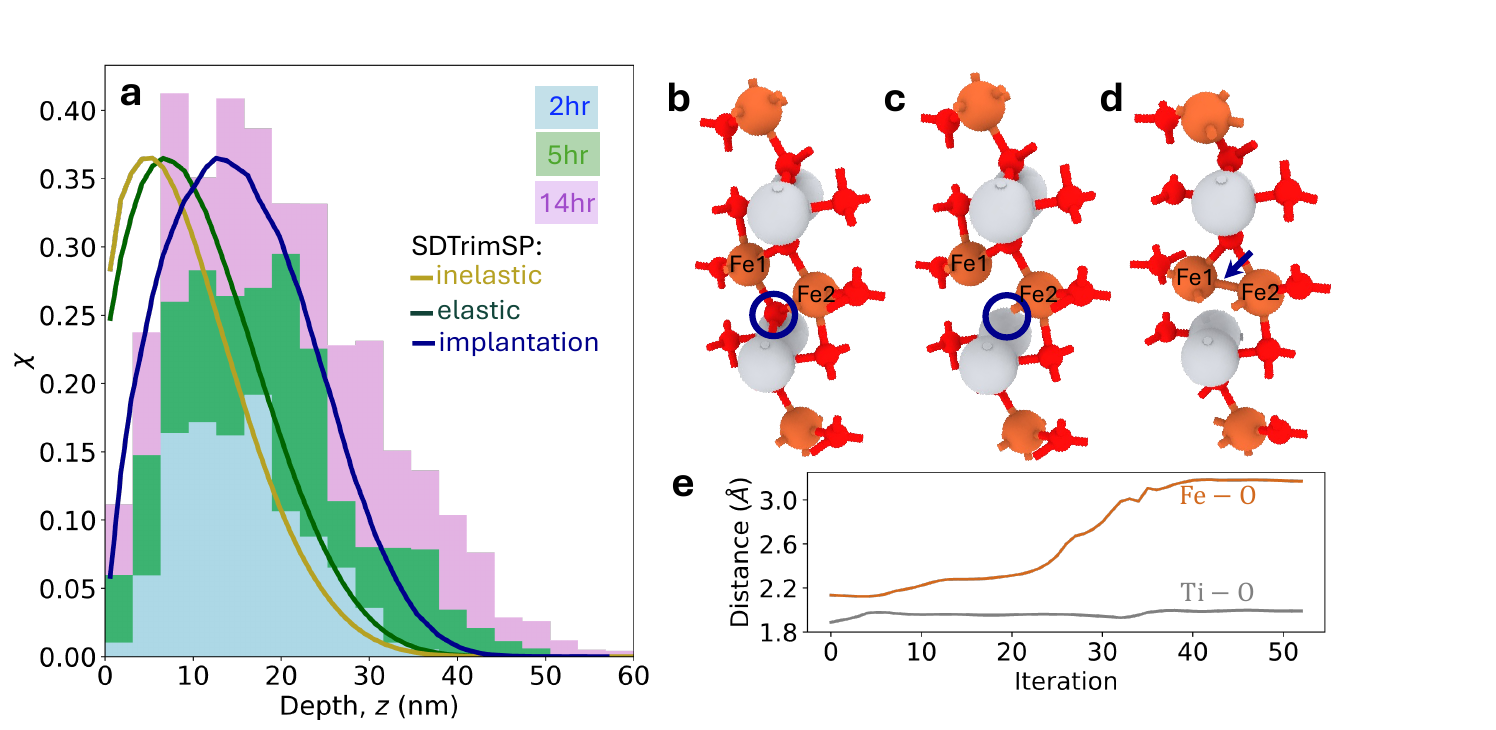}
\caption{Energy deposition and \npFe initiation for \ce{D+}-irradiated ilmenite. (a) Comparison of \npFe distributions ($\chi(z)$) for $\SI{2}{hr}$ (blue), $\SI{5}{hr}$ (green), and $\SI{14}{hr}$ (purple) with SDTrimSP results. Energy losses to electronic (inelastic, yellow) and nuclear (elastic, green) processes from SDTrimSP are scaled to $\chi^{\mathrm{max}}$ for the $\SI{14}{hr}$ exposure to show the depth dependence of energy loss. The implantation profile of \ce{D^+} ions (blue) is also overlaid. (b)-(e) DFT simulation of \ce{FeTiO3}. Snapshots showing (b) starting ilmenite lattice, (c) the initial state with the removal of a single \ce{O} atom (blue circle), and (d) the final state resulting in the formation of an \ce{Fe-Fe} bond (blue arrow). (e) \ce{Fe-O} (gold) and \ce{Ti-O} (gray) distances vs. DFT iteration step. }
\label{fig:simulations}
\end{figure*}

\subsection{\texorpdfstring{\npFe Formation Mechanisms}{npFe Formation Mechanisms}}

The underlying mechanism driving the formation of small \npFe particles ($\SI{<10}{nm}$) from SW irradiation is inherently linked to the generation of point defects. We investigate the role of SW ions in ilmenite to ascertain the potential defect-generating mechanisms leading to the formation of \npFe.

\subsubsection{Mechanisms of Defect Production}\label{subsubsec:defect_production}

As ions traverse a material, they lose energy through collisions, eventually stopping within the material (or transmitting through it). We consider three potential defect-forming mechanisms: (1) energy loss due to inelastic collisions (electronic processes), potentially leading to vacancies formed by electron-excitation-induced Coulombic repulsion \citep{Knotek_1979_auger}, (2) energy loss due to elastic collisions (nuclear processes) with nuclei, creating vacancies by knock-on events,
and (3) the implantation depth for the incident \ce{D^+} ions, potentially resulting in a redox reaction between the implanted ion and \ce{O} atoms \citep{Kuhlman_2015_simulation,Shen_2024_separate}. The SDTrimSP code \citep{mutzke2024sdtrimsp} is used to calculate the electronic/nuclear energy loss, linking the loss to the three defect-forming mechanisms.

Figure~\ref{fig:simulations}a shows a comparison between calculated SDTrimSP distributions (see Section~\ref{subsubsec:SRIM} and Appendix~\ref{app:SDTrimSP}), scaled to the experimental \npFe distributions ($\SI{2}{hr}$ in blue, $\SI{5}{hr}$ in green, $\SI{14}{hr}$ in purple). The SDTrimSP profiles can be used to parse the relative contributions of the three potential defect-forming mechanisms. In Figure~\ref{fig:simulations}a, the agreement between the \ce{D^+} implantation profile and \npFe is remarkable. However, the agreement between implanted \ce{D^+} ions and $\chi(z)$ may not be as informative, since inert ions have also been linked to the formation of \npFe \citep{brunetto2005elastic}. While the implanted \ce{D^+} profile may play a role in the formation of \npFe, it is not the only option, since atomic vacancies are created along the deuterium trajectories through energy-loss mechanisms.
The loss to electronic processes in Figure~\ref{fig:simulations}a (yellow, scaled to $\chi^{\mathrm{max}}$ for $\SI{14}{hr}$ distribution) and nuclear processes (green, scaled to $\chi^{\mathrm{max}}$) have similar profiles, where their separation is comparable to the bin width of $\chi(z)$ (representative of our horizontal error; see Section~\ref{subsubsec: STEM}). Given this experimental uncertainty in $\chi(z)$, we cannot conclusively determine whether electronic or nuclear stopping is more effective in creating \npFe, and the relative importance of these energy-loss/vacancy-creation mechanisms is difficult to distinguish in the ion energy range studied here. However, additional experiments at higher ion energies, where the difference in inelastic and elastic energy-loss processes is larger, could address this question.
Regardless of the dominant defect formation mechanism, atomic vacancies will form along the trajectory of the implanted \ce{D^+} ions. Based on the stoichiometry of ilmenite and efficient momentum transfer from Rutherford scattering, these defects will predominantly be oxygen vacancies (see Appendix~\ref{app:SDTrimSP}).

\subsubsection{Relaxation at Oxygen Vacancy: {Fe-Fe} Bond Formation}

Our DFT calculations explore the scenario of introducing a single oxygen vacancy to determine whether this defect could nucleate \npFe.

Starting from a pristine \ce{FeTiO3} unit cell (Figure~\ref{fig:simulations}b), we performed a structural optimization ($10^{-6}$ energy convergence) yielding an \ce{Fe-O} bond distance of 2.08~\AA{} and a \ce{Ti-O} bond distance of 1.87~\AA{}, both in excellent agreement with literature values \citep{wilson_2005}.
Using a calculational supercell of $2\times 2$ unit cells, we removed an oxygen atom to generate a defect structure (Figure~\ref{fig:simulations}c).
The vacancy formation energy for this defect is $\SI{2.10}{eV}$, matching well with literature values \citep{Luo_2024_vacancies_ilmenite}. Vacancy formation energies were also calculated for Fe (1.70 eV) and Ti (8.0 eV) (see Appendix~\ref{app:vacancy_calc}).
Structural relaxation revealed lengthening of the \ce{Fe-O} bond distance from 2.13~\AA{} to 3.17~\AA{} for the iron atom neighboring the oxygen vacancy (Figure~\ref{fig:simulations}e), indicating a breakage of the \ce{Fe-O} bond. There is also a pronounced shortening of the distance between two adjacent iron atoms (Fe-1 and Fe-2) near the oxygen vacancy, decreasing from an initial 3.05--2.43~\AA{} in intermediate steps before stabilizing at 2.4303~\AA{} in the fully optimized configuration. This is quite close to the bond distance found in $\alpha$-\ce{Fe} (Figure~\ref{fig:SI-DFT_Fe_Ti_distances}b). The removal of an oxygen atom reduces the oxidation state of the \ce{Fe} atoms, altering the electronic environment and promoting the formation of direct Fe--Fe metallic bonds—a configuration absent in the stoichiometric, defect-free structure. The structural rearrangement is energetically favorable, with a total energy reduction of $0.04392\SI{}{Ry}$ ($\sim0.59756\SI{}{eV}$) between the initial and final states, underscoring a strong driving force for local atomic reorganization around the oxygen vacancy.
In contrast, although neighboring \ce{Ti} atoms move closer together (from 3.0503 to 2.8009~\AA{}; Figure~\ref{fig:SI-DFT_Fe_Ti_distances}b), the \ce{Ti-O} bond distance does not change significantly (Figure~\ref{fig:simulations}e and Figure~\ref{fig:SI-DFT_Fe_Ti_distances}a), indicating that \ce{Ti} is not reduced by the oxygen removal.

\section{Conclusion}

Our work highlights the importance of the SW plasma in space-weathering modifications of an important lunar mineral analog: ilmenite. These experiments create \npFe in ilmenite at $\SI{300}{K}$ without the need for other space weathering contributors, confirming the assertion that SW plasma is sufficient to form \npFe in the lunar regolith \citep{Christofferson_1996_ilmenite,Shen_2024_separate}. Additionally, our results reveal systematic trends in the \npFe density versus depth, yet the size of SSW-created \npFe in ilmenite saturates to a relatively constant diameter of $\sim\SI{3}{nm}$. From these results, an independent measure of the surface exposure age for lunar soils is created, which we apply to Apollo soil 71501. Furthermore, our comparison of the measured depth distributions with models of energy loss by SSW particles implicates oxygen vacancies as the critical first step in creating \npFe---a hypothesis confirmed by our DFT calculations. The distinct lack of large \npFe particles in our measurements strongly implicates other formation mechanisms (such as impact events) in those cases. By extending our systematic SSW experiments to other lunar analogs, in tandem with macroscopic optical spectroscopies, new insights into space weathering across the lunar surface could be extracted from remote optical mapping. Similar laboratory methods could be valuable for the interpretation of remote-sensing data from airless bodies throughout the solar system.

\section*{Acknowledgments}
Unprocessed data and relevant codes are publicly available
on Zenodo (doi:10.5281/zenodo.16803752). This work was
directly supported by the NASA solar system Exploration Research Virtual Institute (SSERVI), under cooperative agreement No. NNH22ZDA020C (CLEVER, grant No. 80NSSC23M022). The sample preparation was performed at the Georgia Tech Institute for Matter and Systems, a member of the National Nanotechnology Coordinated Infrastructure (NNCI), which is supported by the National Science Foundation (ECCS-2025462). The authors thank Dr. Karl Lang for supplying the terrestrial ilmenite samples used in this study and Dr. Herbert Funsten for a detailed discussion of SRIM/TRIM. \\

\textit{Software}: Gwyddion \citep{gwyddion},
Hyperspy \citep{delapena2025hyperspy}, SRIM/TRIM \citep{ziegler1977stopping}, pySRIM \citep{Ostrouchov2018}, SDTrimSP \citep{mutzke2024sdtrimsp}, Quantum Espresso \citep{QE_ref}.

\appendix
\renewcommand{\thefigure}{A\arabic{figure}}
\setcounter{figure}{0}

\section{Supplementary Figures}\label{app:SI-figures}

\begin{figure*}[!ht]
\centering
\includegraphics[width=0.9\textwidth]{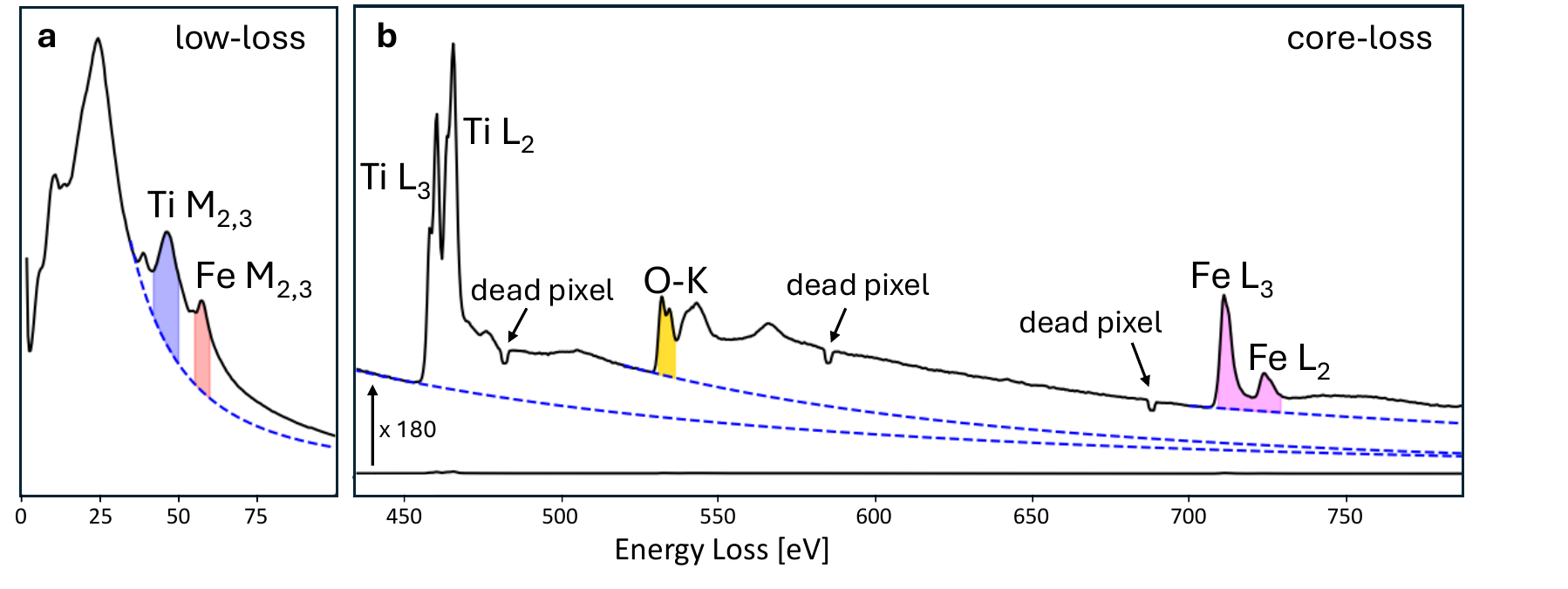}
\caption{Spectral analysis of EELS low-loss and core-loss spectra. EELS low-loss (a) and core-loss (b) spectra from the 2 hr exposed ilmenite sample. Power-law background subtractions (blue dashed lines) are shown for each elemental edge in (a) and (b). (a) Integration window for the EELS \ce{Ti} M-edge (blue) and \ce{Fe} M-edge (red). (b) Integration window for the \ce{O} K-edge (yellow) and \ce{Fe} L-edge (magenta). Dead pixels are due to the gaps between the three EELS detectors (see Section~\ref{sec:methods}).}
\label{fig:SI-EELS_raw}
\end{figure*}

\begin{figure*}[!th]
\centering
\includegraphics[width=0.9\textwidth]{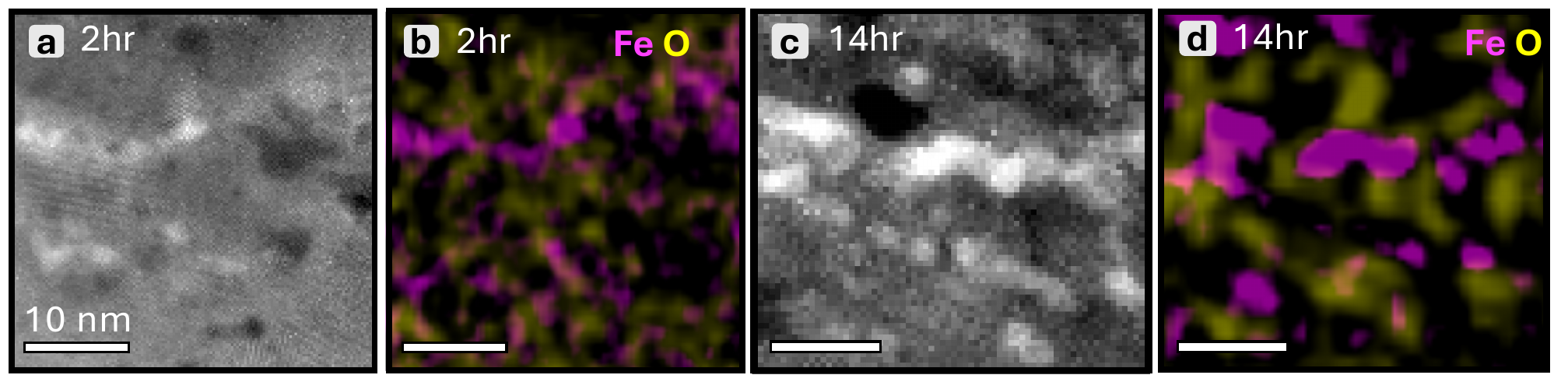}
\caption{Concentration of Fe and O in SSW rims. (a) HAADF image of the $\SI{2}{hr}$ sample. (b) Overlay of \ce{Fe} (pink) and \ce{O} (yellow) concentration maps (Methods) for the $\SI{2}{hr}$ sample. (c), (d) Same as (a), (b), but for the $\SI{14}{hr}$ sample.}
\label{fig:SI-Fe_O_conc}
\end{figure*}

\begin{figure*}[!ht]
\centering
\includegraphics[width=0.9\textwidth]{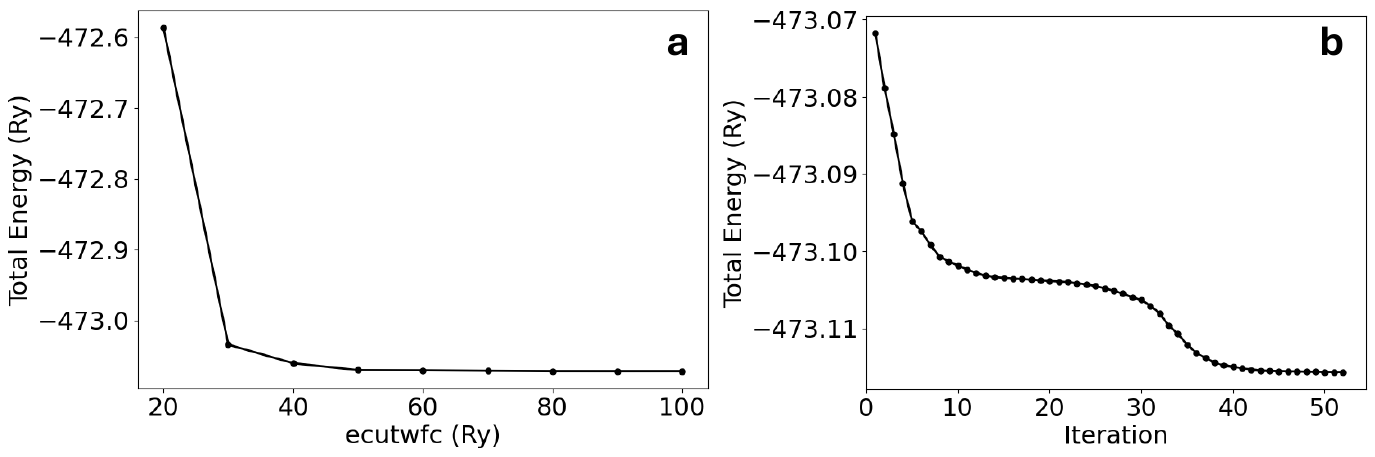}
\caption{DFT evolution of the total energy. (a) Total energy (in Ry) as a function of the plane-wave cutoff energy used in self-consistent field calculation. (b) Evolution of the total energy (in Ry) as a function of geometry optimization iteration steps using 60 Ry as the cutoff energy.}
\label{fig:SI-dft}
\end{figure*}

\begin{figure*}[!ht]
\centering
\includegraphics[width=0.9\textwidth]{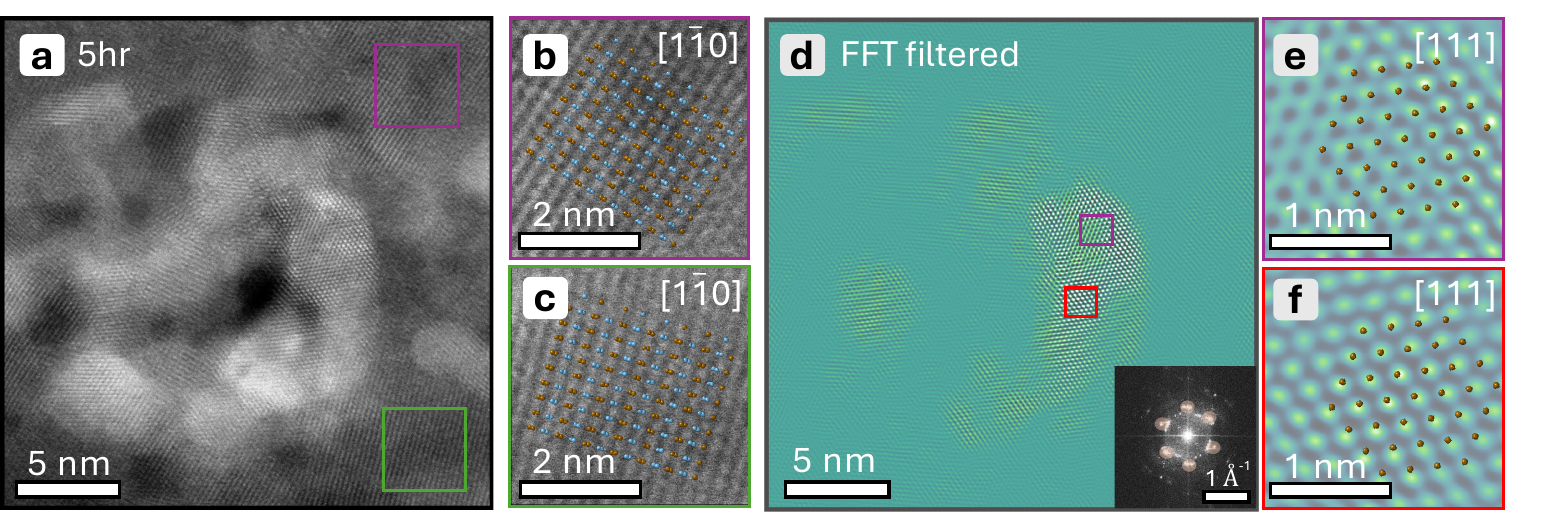}
\caption{Coexistence of crystalline \npFe and ilmenite in the $\SI{5}{hr}$ sample. (a) HAADF image of a \npFe region in the $\SI{5}{hr}$ sample. (b), (c) Zoomed-in views of selected regions in (a), showing crystalline ilmenite within SSW rims (antiamorphous regions). HAADF images in (b) and (c) are overlaid with the ilmenite lattice along the [1$\overline{1}$0] zone axis. (d) Fast Fourier transform (FFT)-filtered image of (a) using  $\alpha$-Fe peaks. Inset: FFT of image (a), indicating a 2.03~\AA~$d$-spacing, consistent with [111] $\alpha$-Fe. (e), (f) Selected regions of crystalline \npFe in (d), overlaid with the $\alpha$-Fe lattice along the [111] zone axis.}
\label{fig:SI-npFe_spacing}
\end{figure*}

\begin{figure*}[!ht]
\centering
\includegraphics[width=0.89\textwidth]{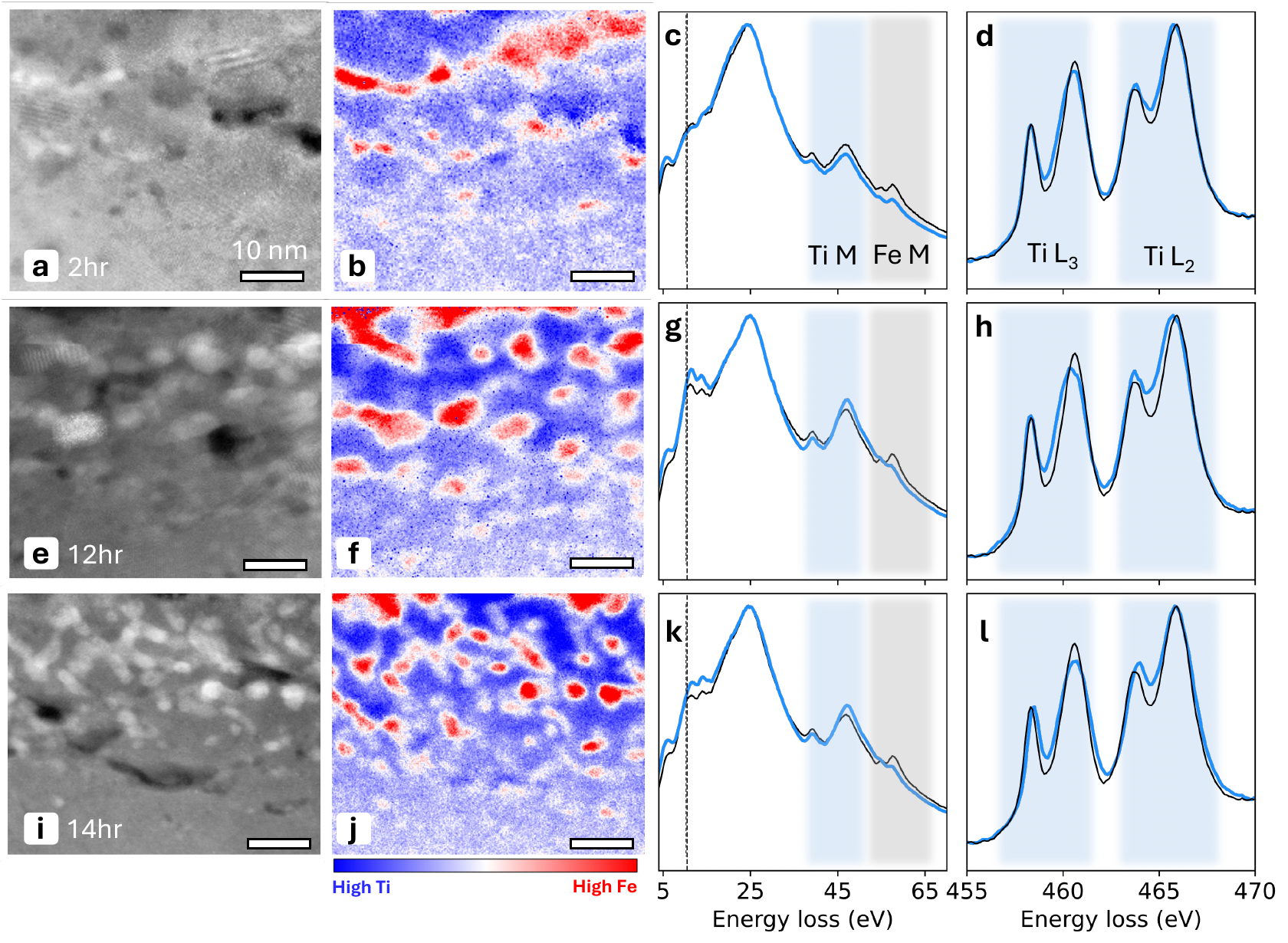}
\caption{EELS characterization of Ti-rich regions in SSW rims. (a) HAADF image of the $\SI{2}{hr}$ sample. (b) \ce{Fe}/\ce{Ti} ratio map (Methods) of the $\SI{2}{hr}$ sample in the same region as (a). (c) EELS low-loss spectra of the \ce{Ti}-rich regions (blue) in (b) along with spectra from unexposed ilmenite (black). Dashed line marks the $\SI{10.5}{eV}$ peak, which arises from the formation of a \ce{TiO2} phase. (d) \ce{Ti} L-edge spectra of the \ce{Ti}-rich regions (blue) alongside an unexposed ilmenite spectrum (black). (e)--(h) Same as (a-d), but for the $\SI{12}{hr}$ sample. (i)--(l) Same as (a-d), but for the $\SI{14}{hr}$ sample. }
\label{fig:SI-Ti-richregions}
\end{figure*}

\begin{figure*}[!ht]
\centering
\includegraphics[width=0.8\textwidth]{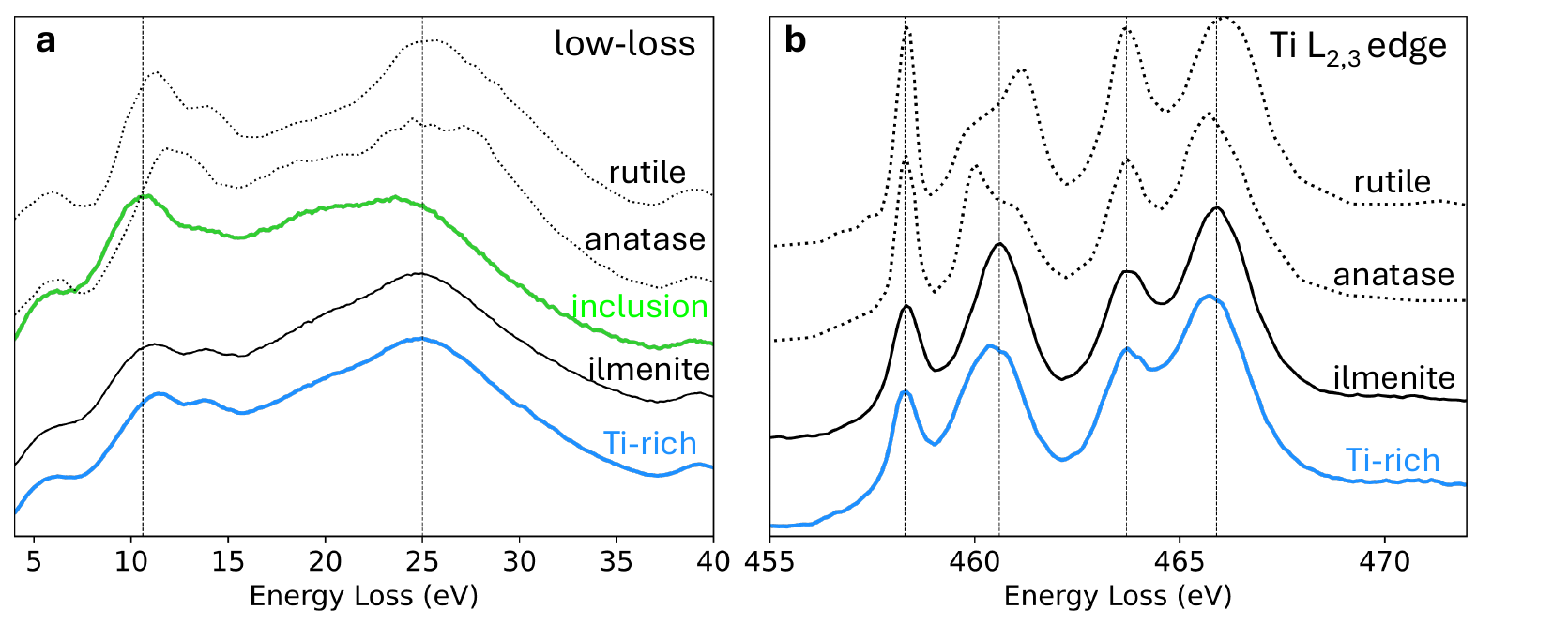}
\caption{Comparison of standards with \ce{Ti}-rich regions from $\SI{12}{hr}$ exposure. (a) EELS low-loss region. (b) \ce{Ti} L$_{2,3}$-edge. Spectra are plotted for dark inclusion (green, averaged over region in Figure~4e), \ce{Ti}-rich region (blue, averaged over region in Figure~4e), ilmenite (black, acquired from unaltered regions), rutile (dotted black; \citealt{Launay_2004_TiO2}), and anatase (dotted black; \citealt{Launay_2004_TiO2}).}
\label{fig:SI-cascade_spectra}
\end{figure*}

\begin{figure*}[!ht]
\centering
\includegraphics[width=0.9\textwidth]{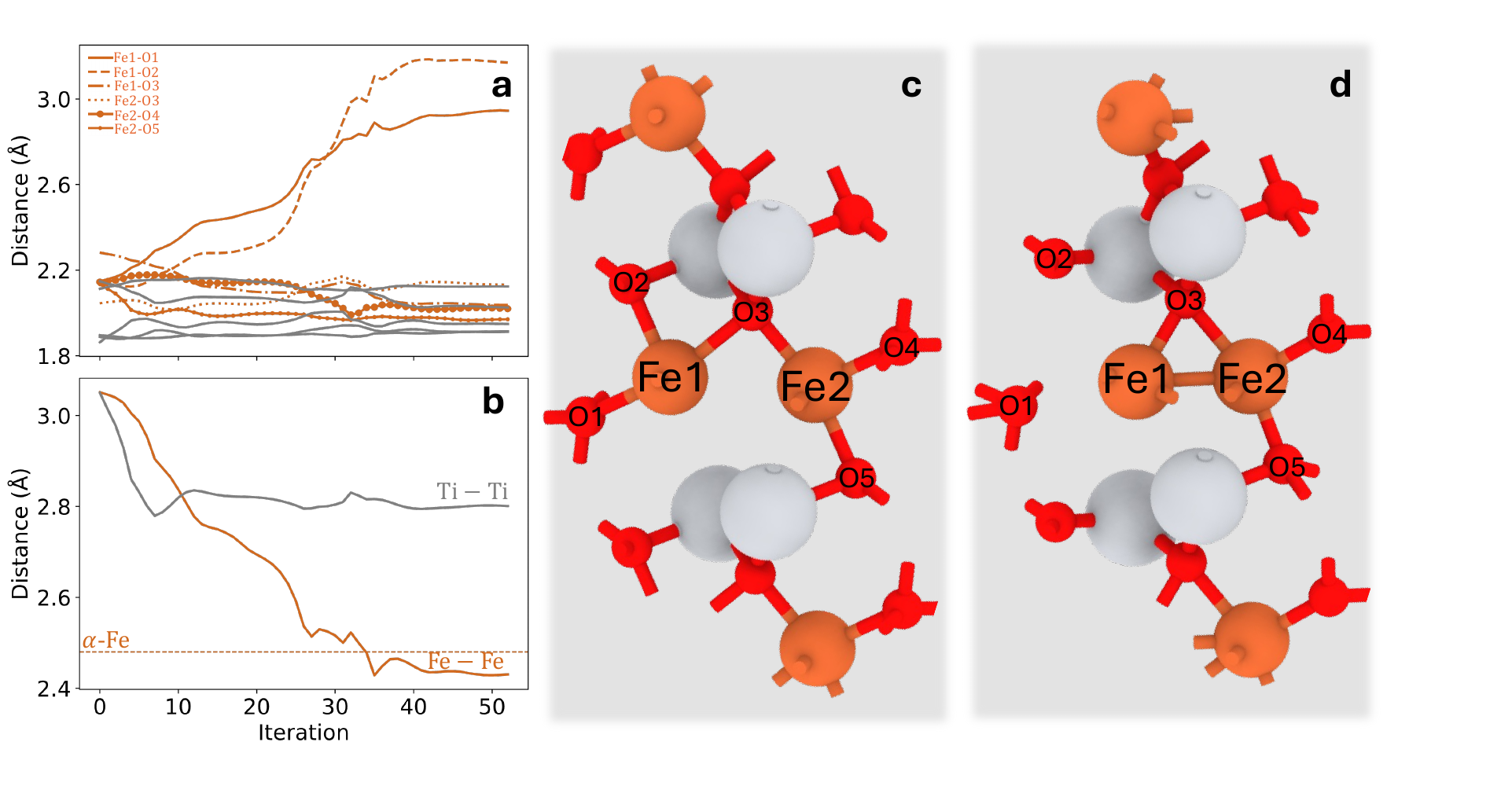}
\caption{Bond distances vs. DFT iteration step. (a) \ce{Fe-O} (gold) and \ce{Ti-O} (gray) bond distances for the iron and titanium atoms neighboring the oxygen vacancy. (b) \ce{Fe-Fe} (gold) and \ce{Ti-Ti} (gray) distances. (c) Atomic model of initial state of ilmenite after removal of the \ce{O} atom with labels indicating the \ce{O} atoms measured in (a). (d) Atomic model of the final (relaxed) state of ilmenite with labels indicating the same \ce{O} atoms as in (c).}
\label{fig:SI-DFT_Fe_Ti_distances}
\end{figure*}

Supplemental Figures shown in the Appendix give information about the raw EELS spectra (Figure~\ref{fig:SI-EELS_raw}), iron and oxygen concentration maps in the space-weathered rim (Figure~\ref{fig:SI-Fe_O_conc}), the energy thresholds and evolution in the DFT calculation (Figure~\ref{fig:SI-dft}), the structural variations in the space-weathered rim (Figure~\ref{fig:SI-npFe_spacing}, Figure~\ref{fig:SI-Ti-richregions}, Figure~\ref{fig:SI-cascade_spectra}), and the evolution of the bonds in the DFT calculation (Figure~\ref{fig:SI-DFT_Fe_Ti_distances}).

\clearpage

\newpage
\section{Comparison of Monte Carlo Codes}\label{app:SDTrimSP}

Both TRIM and SDTrimSP track the trajectories of incident ions as well as subsequent recoil atoms. SDTrimSP extends the framework beyond TRIM by allowing users to specify the interatomic potential and inelastic loss model, addressing some of the systematic errors in SRIM for low-energy ions \citep{Funsten_2001,WITTMAACK201657,Wittmaack_2017}. Another key advancement is its improved models and dynamic treatment of sputtering yield \citep{SZABO202247, Morrissey_2023, mutzke2024sdtrimsp}, although this aspect is not relevant for our work.

\begin{figure*}[!th]
\centering
\includegraphics[width=0.9\textwidth]{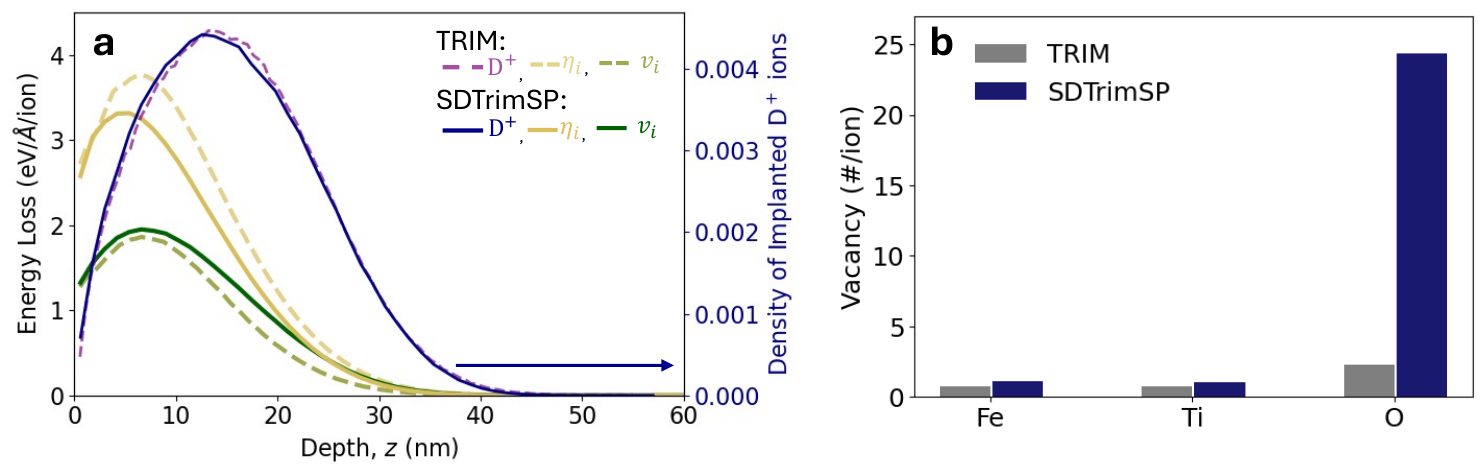}
\caption{Comparison between TRIM and SDTrimSP for $\SI{1}{keV}$ \ce{D+} ions in ilmenite. (a) Energy loss (left axis, $\SI{}{eV/\angstrom/ion}$) to nuclear (green) and electronic processes (yellow) for TRIM (dashed lines) and SDTrimSP (solid lines). The implantation profile (purple) is also shown (right axis, density of implanted \ce{D+} ions) for both codes. (b) Predicted elemental vacancies using the default displacement threshold from TRIM (gray: \ce{Fe}, $\SI{25}{eV}$; \ce{Ti}, $\SI{25}{eV}$; \ce{O}, $\SI{28}{eV}$) and SDTrimSP (blue: \ce{Fe}, $\SI{17}{eV}$; \ce{Ti}, $\SI{19}{eV}$; \ce{O}, $\SI{5}{eV}$).}
\label{fig:SI-srim}
\end{figure*}

Figure~\ref{fig:SI-srim}a compares the energy loss and implantation profiles from TRIM and SDTrimSP. The shapes of the profiles are nearly identical between the two codes, but the magnitude of the energy loss differs, reflecting the improved potentials and inelastic models in SDTrimSP \citep{Wittmaack_2017}. The increased energy loss to nuclear processes in SDTrimSP (green line, Figure~\ref{fig:SI-srim}a) shifts the loss profile in the correct direction, based on the systematic errors of semi-empirical stopping powers from SRIM \citep{Funsten_2001, Helmut_2010, Helmut_2013}. Nevertheless, nuclear processes play a dominant role in the formation of atomic vacancies.
Given the ilmenite stoichiometry and the efficiency of momentum transfer by Rutherford scattering, \ce{O} vacancies are expected to be the most abundant type of atomic vacancies. However, there are inherent uncertainties in the vacancy populations predicted by these MC codes, because vacancies depend on the atom-specific displacement energy ($E_d$), where a displacement occurs when the energy transferred to a recoil atom exceeds $E_d$. To illustrate this uncertainty, Figure~\ref{fig:SI-srim}b shows the predicted number of elemental vacancies per incident ion using the default $E_d$ values from TRIM (gray: \ce{Fe}, $\SI{25}{eV}$; \ce{Ti}, $\SI{25}{eV}$; \ce{O}, $\SI{28}{eV}$) and SDTrimSP (blue: \ce{Fe}, $\SI{17}{eV}$; \ce{Ti}, $\SI{19}{eV}$; \ce{O}, $\SI{5}{eV}$). Because there are no exact values of $E_d$, it is not possible to quantify precisely how much more abundant \ce{O} vacancies are relative to \ce{Ti} or \ce{Fe} vacancies. Nevertheless, fundamental physical reasoning indicates that \ce{O} vacancies will be the most abundant type of vacancy.

\section{Direct Amorphization Models}\label{app:amorphization-models}

The SSW fluence dependence of the peak distribution $\chi^{\mathrm{max}}$ of nanophase iron (\npFe) and the corresponding normalized integrated density $I_\chi$ are shown in Figure~5b. These results are compared with ion-beam-induced amorphization models of well-studied materials (i.e., silicon and quartz wafers, \citep{Dennis_1978_model,Harbsmeier_1998_amorphization}) to assess whether the physics behind these models could serve as a proxy for SSW-induced \npFe formation. The simplest amorphization model is the direct-impact amorphization model, which assumes that every implanted ion produces an amorphous cluster, resulting in a right-circular cylinder with a fixed length and cross-sectional area $A_i$. The solution for the total amorphous area $A_A$ as a function of fluence $\phi$ is \citep{Gibbons_1972}
\begin{equation}
A_A=A_0 \left( 1-e^{-A_i\phi} \right),\label{eq:direct}
\end{equation}
where $A_0$ is the exposed area. For the case of light ions, the formation of an amorphous area requires the overlap of multiple clusters. This effect can be accounted for by summing the exponential terms \citep{Gibbons_1972}:

\begin{equation}
A_A=A_0\left[1-\left(\sum_{k=0}^n\frac{(A_i\phi)^k}{k!}e^{-A_i\phi}\right)\right],\label{eq:overlapping_clusters}
\end{equation}
where $n$ is the number of overlapping clusters. First-order approximations of Equation~\ref{eq:overlapping_clusters} can be described by power-law distributions, resulting in $A_A/A_0=\left(A_i\phi\right)^{n+1}$. For the direct-impact amorphization model ($n=0$), this reduces to $A_A/A_0=\phi/\phi_C$, where $\phi_C=1/A_i$ is the critical fluence. In our case, the \npFe distribution $\chi(z)$ (Figure~5b) is one-dimensional, leading to: $\chi^{max}=\sqrt{A_A/A_0}=\left(\phi/\phi_C\right)^{1/2}$.

\section{Vacancy Formation Energy}\label{app:vacancy_calc}

The vacancy formation energy can be calculated using the following equation:

\begin{equation}
E_f = E_{\mathrm{defect}} - E_{\mathrm{pristine}} + \mu_{\mathrm{element}},
\end{equation}

where $E_{\mathrm{pristine}}$ is the total energy of the pristine system, $E_{\mathrm{defect}}$ is the total energy of the system containing the vacancy, and $\mu_{\mathrm{element}}$ is the chemical potential of the reservoir in which the removed atom is deposited. Consistent with similar calculations (e.g., Luo et al. (2024)), we use a reservoir of elemental \ce{Ti}, \ce{Fe}, and molecular \ce{O2}, to obtain vacancy formation energies of $E_f=\SI{2.10}{eV}$ for \ce{O}, $E_f=\SI{1.70}{eV}$ for \ce{Fe}, and $E_f=\SI{8.0}{eV}$ for \ce{Ti}.

\bibliography{bibliography}
\bibliographystyle{aasjournalv7}

\end{document}